\documentclass[aps,epsfig,nofootinbib]{revtex4}
\pdfoutput=1
\usepackage{graphicx}
\usepackage{dcolumn}
\usepackage{amsmath}
\usepackage{amssymb}
\usepackage{epsfig}
\usepackage[colorlinks,linkcolor=blue,citecolor=blue,urlcolor=blue]{hyperref}

 \newcommand{\bq}{\begin{equation}}
 \newcommand{\eq}{\end{equation}}
 \newcommand{\bqn}{\begin{eqnarray}}
 \newcommand{\eqn}{\end{eqnarray}}
 \newcommand{\nb}{\nonumber}

\begin{document}

\title{Background dynamics of pre-inflationary scenario in Brans-Dicke loop quantum cosmology}
\author{Manabendra Sharma$^1$ \footnote{E-mail address: manabendra@zjut.edu.cn}}
\author{Tao Zhu$^1$ \footnote{ E-mail address: zhut05@zjut.edu.cn}}
\author{Anzhong Wang$^{1,2}$ \footnote{E-mail address: Anzhong$\_$Wang@baylor.edu}}
\affiliation{$^{1}$Institute for Theoretical Physics $\&$ Cosmology,
Zhejiang University of Technology, Hangzhou, 310023, China\\
$^2$GCAP-CASPER, Department of Physics, Baylor University, Waco, TX, 76798-7316, USA }

\date{\today}

\begin{abstract}

Recently the background independent nonperturbative quantization   has been  extended to various theories of gravity  and the corresponding quantum  
effective cosmology has been derived, which provides us with necessary avenue to explore the pre-inflationary  dynamics. Brans-Dicke (BD) loop 
quantum cosmology (LQC) is one of such theories whose effective background dynamics is considered in this article. Starting with a quantum 
bounce,  we explore the pre-inflationary dynamics of a universe sourced by a scalar field with the Starobinsky potential in BD-LQC. Our study is 
based on the idea that though Einstein's and  Jordan's frames are  classically equivalent up to a conformal transformation  in BD theory, this  is 
no longer true after quantization. Taking the Jordan frame as the physical one we explore in detail the bouncing scenario which is followed by a 
phase of a slow roll inflation. The three phases of the evolution of the universe,  namely, {\em bouncing, transition from quantum bounce to 
classical universe, and the  slow roll inflation}, are noted for an initially kinetic energy dominated bounce.  In addition, to be consistent with 
observations, we also identify  the allowed phase space of initial conditions that would produce at least 60 e-folds of expansion during the slow 
roll inflation.

\end{abstract}

\maketitle
 
\section{Introduction}\label{Introduction}

The inflationary cosmology as proposed by Alan Guth in  1981  \cite{Guth} provides a solution to some of the puzzles of the Standard Big Bang Cosmology by introducing an epoch of nearly exponential expansion in the early universe. Other than serving as a mechanism to solve problems like horizon, flatness and entropy etc., it is much more powerful in predicting the cosmos we see today  \cite{Realized}. It offers a first ever causal explanation for the origin of inhomogeneities in the universe. To be precise, it predicts the primordial power spectra whose evolutions explain both the formation of the large scale structure and small inhomogeneities present in the cosmic microwave background (CMB)  \cite{Guth,CMB}. Today, the unprecedented success of the inflationary paradigm is based on observational precisions  \cite{Success1}-\cite{Success3}. Though very successful, the paradigm still suffers from  issues  like initial singularity and trans-Planckian problem. For a review on conceptual issues of inflationary cosmology see  \cite{UC}.

For every expanding Friedmann-Lametre-Robertson-Walker (FLRW) universe, there always appears  a big-bang singularity  in general relativity (GR), if matter satisfies certain energy conditions  \cite{HE73}. This strong curvature singularity at which physics comes to halt is an artifact, because we are pushing the classical theory of gravity to a region where it is no longer valid. Mathematically speaking, the geodesics are past incomplete and hence the affine parameter cannot be extended to infinite past leading to a focusing of congruence of geodesics \cite{Raychaudhuri,LargeScaleStructure}. This causes the breakdown of the notion of space-time itself. The result of focusing theorem is valid both for null and time like geodesic \cite{EricPoisson}. It can be shown that this issue is independent of the symmetry of the metric considered unless and until certain energy conditions are hold good and depends solely on the average value of the Hubble parameter \cite{MyThesis1}. Neither Standard model of cosmology nor inflationary paradigm have a say on this issue. An alternative scenario, called Bounce, within the framework of classical theory of gravity requires either the consideration of exotic stress-energy tensor or the modification of the gravity sector of the Einstein-Hilbert action which is evident from the Raychaudhuri equation in the absence of the centrifugal term  \cite{Raychaudhuri}. In addition to solving the puzzles of the standard model of cosmology  \cite{SolutionHPFP}, these classical bouncing scenarios involve circumvention of the singularity by assuming the universe to start from a contracting phase and then bouncing back to an expanding phase well before reaching the Planck length  \cite{MyThesis1}. However these scenarios are also faced with serious setbacks like instability problem  \cite{Instabilities}. For a review on classical bouncing scenarios and their drawbacks see  \cite{ReviewOnBouncing1,ReviewOnBouncing2}. Amongst all the drawbacks the fact that the universe becomes extremely dense early on suggests that quantum gravitational effects would become important at such small length and one must adopt quantum gravity to describe the ``zone of ignorance." Therefore, it is natural to speculate that in a true theory of quantum gravity quantum mechanics would intervene to avert the singularity. 

Loop quantum gravity (LQG) is a candidate of quantum gravitational theory where it takes the premise of gravity as a manifestation of geometry of space-time seriously and then systematically constructs a specific theory of quantum Riemanian geometry. For a review we refer the reader to   \cite{Thiemann1}. LQG stands out as a leading non-perturbative and background-independent approach to quantize gravity \cite{QuantumGravity1}. At its depth, this theory brings out a fundamental discreteness at the Planck scale wherein the underlying geometric observables, such as areas of physical surfaces and volumes of physical regions,  are discrete in nature  \cite{Asthekar1,Asthekar2,Bianchi1}. The avenue of loop quantum cosmology (LQC) is an application of LQG techniques to the symmetry reduced space-time, homogeneous space-time for cosmology in particular \cite{Bojowald1}. In LQC the singularity is resolved in the sense that physical macroscopic observables, such as energy density and curvature which diverge at the big bang in GR, have a finite upper bound. This finiteness of the macroscopic parameters above owes to the fact that they have a dependence on the microscopic parameter of the theory called the fundamental area gap whose smallest eigenvalue is nonzero  \cite{Asthekar3,Asthekar4,Asthekar5}. Thus, a contracting FLRW universe would bounce back to an expanding one avoiding the formulation of singularity in LQC as there is a maximal value of energy density. This is achieved without adding any nontrivial piece of matter unlike in the case of classical bounce. This quantum bounce occurs purely due to quantum geometric effects which act as a novel repulsive force that can be easily seen from the quantum corrected Friedmann and Raychaudhuri equations to be discussed in Sec.\ref{Cosmology}. Also, it is to be noted that in all the different classes of space-times permitting different sets of symmetries  \cite{WE}, including Bianchi and Gowdy models, the singularity is resolved in the framework of LQC \cite{Asthekar6}-\cite{Brizuela2}. Based on the loop quantization of Brans-Dick (BD) theory the dynamics of BD-LQC has also been explored  \cite{Jin}. For a review of singularity resolution in LQC,  see  \cite{Asthekar9,Singh3}.

In addition to the above, the fact that the universe must have expanded at least 50 e-folds so as to be consistent with the current observations leads to a problematic situation if the universe had expanded a little more than 70 e-folds. In fact, had it been so (which is true for a large class of inflationary models  \cite{5}), it turns out that the wavelengths of all fluctuation modes which are currently inside the Hubble radius were smaller than the Planck length at the onset of the   inflation. This is coined as the trans-Planckian issue in \cite{6}, which challenges the validity of the assumption that matter fields are quantum in nature but spacetimes can still be treated classically which are used at the beginning of inflation in order to make predictions  \cite{CMB}. Thus, once again it calls for quantum treatment of spacetimes. Moreover, the inflationary paradigm usually sets the adiabatic  vacuum state at the time when the wavelengths of fluctuations were well within the Hubble horizon during the phase of inflation. This treatment, however, allows the ignorance of the dynamics prior to the onset of the inflation, even when the modes were well inside the Hubble horizon. For more details regarding the sensitivity of the inflationary dynamics to quantum gravitational effects we refer readers to  \cite{6,8, Zhu:2013fha, Zhu:2015xsa, UnivFeatB}. 

As cited above all these issues motivate to look for quantum gravity candidates and LQC stands out as, with robustness, a competing one that replaces the singularity by a quantum bounce which is followed by a desired slow roll inflation  \cite{18}. Now LQC is in a position to undergo experimental tests and to look for observational signatures of quantum bounce, pre-inflationary dynamics in current/forth-coming observations. In fact three major streams of calculations of cosmological perturbations, namely, \textit{dressed metric}  \cite{Asthekar9}, \textit{deformed algebra}  \cite{14}, and \textit{hybrid approaches} \cite{Mendez} have been carried out and studied numerically and analytically in  \cite{10}-\cite{navascures_hybrid_2016}. It has been reported that the \textit{deformed algebra} approach is already inconsistent with current observations  \cite{21,arxiv}.

Coming back to the inflationary paradigm, this standard inflationary scenario is based on a canonical scalar field in the framework of GR. Viability of different models of inflation in the framework of standard inflationary scenario have been extensively studied  \cite{13JCAP}-\cite{17JCAP}. An important category of inflationary models are based on the extended theories of gravity. This classical inflationary models on the extended theories of gravity have been extensively studied in literature  \cite{18JCAP}-\cite{46JCAP}. In particular the study of inflation in BD gravity has been carried out in detail in  \cite{JCAP}. Wherein the inflationary observables containing the scalar spectral index, the tensor to scalar ratio, the running of the spectral index and the equilateral non-Gaussianity parameters in terms of general form of potential in the Jordan frame are obtained and results are compared in the light of Planck 2015 data  \cite{JCAP}. However as the early universe is extremely hot, it is also important to consider the quantum gravitational effects of spacetime in the early universe. 
Now as mentioned above for the last two and half decades LQG has caught attention of the community and  been widely investigated  \cite{QuantumGravity1, 43ZhuTao}, as it gives a background independent way to quantize GR. In this framework of LQG, it is remarkable that GR can be non-perturbatively quantized. 
Recently this promising loop quantization technique has been extended to theories of modified gravity, for instances, BD theory  \cite{35ZhuTao,36ZhuTao}, metric $f(R)$ theories  \cite{37ZhuTao,38ZhuTao} and scalar-tensor theories  \cite{17ofZhaAndMa}. Thus, it opens up an interesting avenue to explore the early universe when geometry is quantized beyond GR. 
 That BD theory in Jordan's frame is  equivalent to Einstein's frame classically up to a conformal transformation is no longer true  after quantization in the two frames \cite{17ofZhaAndMa}.
In \cite{Jin} the  pre-inflationary dynamics of BD theory is extensively studied by taking Einstein's frame as the physical one.

In this paper we focus on studying the effective BD-LQC by taking Jordan's frame as the physical one for a universe sourced by a scalar field with the Starobinsky potential. This paper is divided broadly into three main sections: Sec.\ref{Cosmology}, Sec.\ref{CosmologyResults} and Sec.\ref{DR}. In Sec.\ref{Cosmology} we present briefly the background cosmology in BD gravity in classical set-up. Wherein we present the equations of motion of background dynamics both in Einstein's and Jordan's frames and emphasize that both are classically equivalent to each other up to a conformal transformation, provided that the same coordinate system  is used. Sec.\ref{CosmologyResults} contains our studies and main results of a loop quantum corrected BD cosmology sourced by a scalar field with the Starobinsky potential. This is further divided into two subsections \ref{EV} and \ref{NA}. In subsection \ref{EV} we set up the stage by presenting the loop quantum corrected equations of motion for the background dynamics of BD theory in Jordan's frame. We analyze the quantum corrected Friedmann and Raychaudhuri equations  in detail to see whether a quantum bounce exists or not. In addition to stating the necessary conditions,  we put forward an effective definition of pressure. In the subsection \ref{NA} we present all our numerical results for a kinetic dominated bouncing scenario. Wherein we check whether a slow roll inflation is attained and subsequently calculate the number of e-folds of expansion generated for different values of the field at the quantum bounce. Finally we discuss and conclude our work in \ref{DR}.

\section{Dynamics of the Background Cosmology}\label{Cosmology}

\subsection{Classical dynamics of slow roll inflation in BD theory}

BD proposed a specific form of the scalar tensor gravity, that is based on the Mach principle, which implies that the inertial mass of an object depends on the matter distribution in the universe  \cite{59JCAP}. The governing action for four dimensional BD theory is given by  \cite{59JCAP},
\begin{equation}
S_J=\int d^4x \sqrt{-g} \left[ \frac{M_{Pl}}{2} \phi R + \frac{M_{Pl}}{\phi}w_{BD}X-V(\phi)\right],
\end{equation}
where $g$ is the determinant of the spacetime metric $g_{\mu \nu}$, $R$ is the four dimensional  Ricci scalar, $\phi$  the BD scalar field, $w_{BD}$  the dimensionless BD parameter, whereas $X$ is defined as $X\equiv -\frac{1}{2} g^{\mu \nu} (\partial_\mu \phi ) (\partial_\nu \phi)$,  and $V(\phi)$ represents the potential of the scalar field $\phi$. In this work we adopt the FLRW metric and in contrary to the original BD theory we introduce the field potential $V(\phi)$. It is to be noted that the reduced Planck mass $M_{Pl}$ is defined as $M_{Pl}^{-2}=8 \pi G$ and the Planck mass $m_{Pl}$ is defined as $m_{Pl}^{-2}=G$. 

Now  $f(R)$ gravity with the action 
\begin{equation}
S= \int d^4 x \sqrt{-g} \frac{M_{Pl}^2}{2} f(R), 
\end{equation}
can be easily obtained from BD theory if we consider the following correspondence,
\begin{equation}
\frac{\phi}{M_{Pl}}=\frac{{\rm d} f}{{\rm d} R},~~V(\phi)= \frac{M_{Pl}^2}{2}\left( R \frac{{\rm d} f}{{\rm d}R}-f \right),~~w_{BD}=0.
\end{equation}
This in turn gives the Starobinsky inflation with the action  \cite{CMB}
\begin{equation}
S_{R^2}= \int d^4x \sqrt{-g} \frac{M_{Pl}^2}{2}\left( R + \frac{R^2}{6M^2} \right), \label{Action}
\end{equation}
as a particular case of BD theory with the following transformation,
\begin{equation}
\frac{\phi}{M_{Pl}}= 1+ \frac{R}{3M^2},~~V(\phi)=\frac{3M^2}{4}(\phi-M_{Pl})^2,~~w_{BD}=0.
\end{equation}
Now to get the equation of motion of the background, we invoke a spatially flat, homogeneous and isotropic metric in the Jordan frame which takes the  form
\begin{equation}
ds^2=-dt^2 + a^2(t) \delta_{ij} dx^{i}dx^{j},
\end{equation}
where $a(t)$ is the scale factor of the universe and  $t$ is the cosmic time defined in Jordan frame. The background dynamics of the universe, can be obtained by varying the action w.r.t the metric $g_{\mu \nu}$ and $\phi$
 \cite{49JCAP,69JCAP} to give the modified Friedmann and Klein-Gordon equations,
\begin{eqnarray}
3 M_{Pl}^2\left( H + \frac{\dot{\phi}}{2 \phi} \right)^2= \frac{M_{Pl}^2 \rho_{\phi}}{\phi^2},\\
\ddot{\phi} + 3H\dot{\phi} + \frac{2}{\beta M_{Pl}}\left[\phi V_{\phi}(\phi) -2 V(\phi)\right]=0,
\end{eqnarray}
where $H\equiv {\dot{a}}/{a}$ is the Hubble parameter, $\beta= 2 w_{BD} +3 $ and $\rho_{\phi}$ is defined as $\rho_{\phi}\equiv ({\beta}/{4}) \dot{\phi}^2 + {\phi V(\phi)}/{M_{Pl}}$  is the effective energy density of the BD scalar field. It should be noticed that for the Starobinsky inflation we have $\beta=3.$

\subsection{Starobinsky potential in Einstein frame}
In this subsection we present, for the sake of completeness, the classical equation of motion of the background universe sourced by the Starobinsky potential in the Einstein's frame. For detailed analysis of Starobinsky inflation in Einstein's frame with loop quantum correcton see  \cite{Jin}. Classically, the Jordan and the Einstein frames are equivalent to each other up to a conformal transformation when the same coordinate system is employed. Thus under a conformal transformation 
\begin{equation}
\hat{g}_{\mu \nu} = \frac{\phi}{M_{Pl}}g_{\mu \nu}, \label{CT}
\end{equation}
the action Eq.(\ref{Action}) can be modeled to that of a minimally coupled scalar field $\chi$ in the Einstein frame where we use a hat to represent quantities in this frame. Thus, under this conformal transformation  the action  gets recasted into 
\begin{equation}
S_E= \int d^4x \sqrt{-\hat{g}} \left[ \frac{M_{Pl}^2}{2}\hat{R}-\frac{1}{2}\hat{g}_{\mu \nu} \partial_{\mu}\chi \partial_\nu \chi - U(\chi) \right],
\end{equation}
and we can easily recognise that 
\begin{eqnarray}
U(\chi)&=& e^{-2 \sqrt{\frac{2}{\beta}}\frac{\chi}{M_{Pl}}}V(\phi),\\
\frac{\phi}{M_{Pl}} &=& e^{\sqrt{\frac{2}{\beta}}\frac{\chi}{M_{Pl}}}.
\end{eqnarray}
Now, for the Starobinsky inflation the correspondence between scalar field $\chi$ and the Ricci scalar is given by 
\begin{equation}
\frac{\chi}{M_{Pl}}= \sqrt{\frac{3}{2}}ln\left( \frac{\phi}{M_{Pl}} \right)=\sqrt{\frac{3}{2}}ln\left( 1+ \frac{R}{3M^2} \right),
\end{equation}
which gives 
\begin{eqnarray}\nonumber
U(\chi) = \frac{3}{4} M^2 M_{Pl}^2 \left( 1- e^{-\sqrt{\frac{2}{\beta}}\frac{\chi^2}{M_{Pl}^2}} \right),\;\;\;
\beta = 3. 
\end{eqnarray}
Note that the Starobinsky potential $U(\chi)$ with $\beta=3$ can be extended to a subclass of E type $\alpha$ attractor inflationary models  \cite{8ZhuTao,9ZhuTao,19JCAP}. 
 Now considering a universe with  the flat FLRW metric in the Einstein frame
\begin{equation}
ds^2= -d\hat{t}^2 + \hat{a}^2 \delta_{ij}d{x}^i d{x}^j, 
\end{equation}
where $\hat{a}(\hat{t})$, $d{x}^i$ being the scale factor and comoving coordinates in Einstein's frame. The background dynamics, in this frame, are given by the modified Friedmann and Klein-Gordon equations
\begin{eqnarray}
\hat{H}^2 = \frac{1}{3 M_{Pl}^2} \hat{\rho}_{\chi}, \\
\frac{{\rm d}^2 \chi}{{\rm d}\hat t^2}+ 3 \hat{H} \frac{{\rm d} \chi}{ {\rm d} \hat t} + U(\chi) &=& 0,
\end{eqnarray}
where the effective energy density is given by $\hat{\rho}_{\chi}= \frac{1}{2} \dot{\chi}^2 + U(\chi)$.

\subsection{Equivalence between the Einstein and the Jordan frame}

As has been stated above the Einstein and Jordan frames are equivalent upto a conformal transformation  at the classical level. Under this transformation (\ref{CT}) the relationship between the corresponding dynamical quantities of the theory in the two frames can be immidiately established to be 
\begin{equation}
\hat{a}(\hat{t}) = \sqrt{\frac{\phi}{M_{Pl}}}a(t), ~~ {\rm d} \hat{t} = \sqrt{\frac{\phi}{M_{Pl}}} {\rm d} t,
\end{equation}
for the scale factors and the cosmic times. With these the following relations can be easily found
\begin{eqnarray}
\hat{H} &=& \sqrt{\frac{M_{Pl}}{\phi}} \left( H + \frac{\dot{\phi}}{2 \phi} \right),\\
\frac{{\rm d}\chi}{{\rm d}\hat{t}} &=& \sqrt{\frac{\beta}{2}} \left( \frac{M_{Pl}}{\phi} \right)^{\frac{3}{2}} \dot{\phi},\\
\frac{{\rm d}^2 \chi}{{\rm d} \hat{t}^2} &=& \sqrt{\frac{\beta}{2}}M_{Pl}^2 \left( \frac{\ddot{\phi}}{\phi}^2 -\frac{3}{2} \frac{\dot{\phi}^2}{\phi^3} \right),\\
\hat{\rho}_{\chi} &=& \left( \frac{M_{Pl}}{\phi} \right)^3 \left[ \frac{\beta}{4} \dot{\phi}^2 + \frac{\phi}{M_{Pl}} V(\phi)  \right].
\end{eqnarray}
Note that though these transformation between the two frames hold good at the classical level, it does not turn out to be so after quantization. In fact  the effective dynamics is different and this point is what we explore in this work. 

\section{Effective background dynamics of Brans-Dicke loop quantum  cosmology}\label{CosmologyResults}

\subsection{Effective background dynamics of BD-LQC in Jordan's frame}\label{EV}

In this subsection we present the required equations to numerically simulate the effective quantum corrected background dynamics in BD-LQC  in Jordan's frame.  The modified Friedmann equation and the Klein-Gordon equation are sufficient to study the background dynamics. The quantum corrected Friedmann equation for the effective BD-LQC theory in Jordan frame is  \cite{36ZhuTao}
\begin{equation}
\left( H + \frac{\dot{\phi}}{2\phi}\right)^2 = \left[ \frac{1}{\phi}\sqrt{\frac{\rho_{\phi}}{3}}\sqrt{1-\frac{\rho_{\phi}}{\rho_c}} + \frac{\dot{\phi}}{2\phi} \left( 1-\sqrt{1-\frac{\rho_{\phi}}{\rho_c}}\right) \right]^2, \label{FD}
\end{equation}
where  the effective energy density  is given by $\rho_{\phi}= \frac{\beta}{4} \dot{\phi}^2 + \frac{\phi V(\phi)}{M_{Pl}}$. The quantum corrected Klein-Gordon equation for BD theory in Jordan's frame is
\begin{equation}
\ddot{\phi} + 3H \dot{\phi} + \frac{2}{\beta M_{Pl}} \phi V_{\phi} + \frac{2}{\beta M_{Pl}} V(\phi)\left( 1- 3 \sqrt{1- \frac{\rho_{\phi}}{\rho_c}} \right)=0. \label{KG}
\end{equation}
As stated above, we need to solve these two equations to explore the background dynamics. Eq.(\ref{FD}) immediately gives 
\begin{equation}
\left( H + \frac{\dot{\phi}}{2\phi}\right) =  \pm \left[ \frac{1}{\phi}\sqrt{\frac{\rho_{\phi}}{3}}\sqrt{1-\frac{\rho_{\phi}}{\rho_c}} + \frac{\dot{\phi}}{2\phi} \left( 1-\sqrt{1-\frac{\rho_{\phi}}{\rho_c}}\right) \right]. \label{FD1}
\end{equation}
Thus we get two solutions. The $``+"$ solution of Eq.(\ref{FD1}) gives 
\begin{equation}
H= \frac{1}{\phi}\left[ \sqrt{\frac{\rho_{\phi}}{3}} -\frac{\dot{\phi}}{2}\right]\sqrt{1-\frac{\rho_{\phi}}{\rho_c}}, \label{H1}
\end{equation}  
and the    $``-"$ solution gives 
\begin{equation}
H = -\frac{\dot{\phi}}{\phi} - \frac{1}{\phi}\left( \sqrt{\frac{\rho_{\phi}}{3}} -\frac{\dot{\phi}}{2}\right) \sqrt{1-\frac{\rho_{\phi}}{\rho_c}}. \label{H2}
\end{equation}
Before defining some useful parameters to carry out qualitative and quantitative analysis of our models considered in this work, let us check if we attain a nonsingular bouncing model in each of theses two cases, namely, given by, Eq.(\ref{H1}) and Eq.(\ref{H2}), respectively. 

A nonsingular bounce is obtained as when an initially contracting universe goes to an expanding one through a minimum in the scale factor but not zero. Since the Hubble parameter is proportional to the fractional rate of change of volume, $H \propto \frac{\triangle V}{V \triangle t}$,  therefore a contracting (expanding) phase is specified by a negative (positive) value of the Hubble parameter H. However this is a necessary but not sufficient condition for bounce to occur. Along with this the slope of the slope of the scale factor must be positive at  the turn around to achieve the minimum and hence the bounce. Mathematically speaking the following two conditions must be met for the bounce to occur,
\begin{eqnarray}
\left(\frac{{\rm d}a}{{\rm d}t}\right)_B=0,\label{Condition1} \\ 
\left( \frac{{\rm d}^2 a}{{\rm d}t^2} \right)_B > 0. \label{Condition2}
\end{eqnarray}
From here onwards the subscript B will be used to denote the point of the bounce. Now let us examine if the two conditions given by Eqs.(\ref{Condition1}) and (\ref{Condition2}) are satisfied for the two solutions of   Eq.(\ref{H1}) (for the $``+"$ solution) and Eq.(\ref{H2}) (for the $``-"$ solution) respectively. In the contracting phase,  as the volume of the universe decreases and therefore the energy density of the universe increases, we expect the universe to turn around when it reaches the maximal energy scale $\rho_c$ given by the theory due to quantum geometric effects. Thus, we need first to check if the Hubble parameter is zero when $\rho_{\phi}= \rho_{c}$. 
From Eq.(\ref{H1}) it is clear that 
\begin{equation}
H=\frac{1}{\phi}\left( \sqrt{\frac{\rho_{\phi}}{3}} -\frac{\dot{\phi}}{2}\right)\sqrt{1-\frac{\rho_{\phi}}{\rho_{c}}}= 0 ~~~at~~ \rho_{\phi}=\rho_{c},
\end{equation}
and for  Eq.(\ref{H2}) we find
\begin{equation}
H=-\frac{\dot{\phi}}{\phi} -\frac{1}{\phi}\left( \sqrt{\frac{\rho_{\phi}}{3}} -\frac{\dot{\phi}}{2}\right)\sqrt{1-\frac{\rho_{\phi}}{\rho_{c}}}= 0 ~~~at~~ \rho_{\phi}=\rho_{c} ~~if ~and~ only~ if ~~\dot{\phi}|_{\rho_{\phi}=\rho_c}.
\end{equation}
But with these, only Eq.(\ref{Condition1})  is fulfilled. We need to see explicitly for each of the solution of H whether the second condition Eq.(\ref{Condition2}) is also satisfied or not (at $\rho_{\phi}=\rho_{c})$.  
As we know that given the Friedmann and Klein-Gordon equations,  the Raychaudhuri equation can be derived using the following identity
\begin{equation}
\frac{\ddot{a}}{a} = H^2 + \frac{{\rm d}H}{{\rm d}t}.
\end{equation}
Now let us derive it for each of the equations (\ref{H1}) and (\ref{H2}) case by case. Considering first Eq.(\ref{H1}), as we have seen above that the Hubble parameter becomes zero as $\rho_{\phi}=\rho_c$, therefore, 
\begin{equation}
\left( \frac{\dot{a}}{a} \right)_{\rho_{\phi}=\rho_{c}} = \left( \frac{{\rm d} H}{{\rm d} t} \right)_{\rho_{\phi}=\rho_{c}}.
\end{equation}
Then we find
\begin{eqnarray}
\frac{{\rm d}H}{{\rm d}t}&=& \frac{{\rm d}}{{\rm d}t} \left[ \frac{1}{\phi}\left[ \frac{\rho_{\phi}}{3} -\frac{\dot{\phi}}{2}\right]\sqrt{1-\frac{\rho_{\phi}}{\rho_c}} \right],\\
 &=& \frac{{\rm d}}{{\rm d}t} \left[ \frac{1}{\phi}\left( \frac{\rho_{\phi}}{3} -\frac{\dot{\phi}}{2}\right)\right]\sqrt{1-\frac{\rho_{\phi}}{\rho_c}} ~+~ \frac{1}{2\phi}\left(\sqrt{\frac{\rho_{\phi}}{3}}-\frac{\dot{\phi}}{2} \right) \frac{-\frac{1}{\rho_c}\frac{{\rm d}\rho_{\phi}}{{\rm d}t} }{\sqrt{1-\frac{\rho_{\phi}}{\rho_{c}}}}.
\end{eqnarray}
Evaluating the above expression at $\rho_{\phi}= \rho_{c}$,  we are left with only the second term 
\begin{equation}
\left(\frac{{\rm d}H}{{\rm d}t}\right)_{\rho_{\phi}=\rho_c}=\left( \frac{1}{2\phi}\left(\sqrt{\frac{\rho_{\phi}}{3}}-\frac{\dot{\phi}}{2} \right) \frac{-\frac{1}{\rho_c}\frac{{\rm d}\rho_{\phi}}{{\rm d}t} }{\sqrt{1-\frac{\rho_{\phi}}{\rho_{c}}}} \right)_{\rho_{\phi}=\rho_{c}}.
\end{equation}
Care must be taken while taking the limit. Note that first we need to evaluate the differentiation and then take the limit. In order to do that let us begin with the definition of $\rho_{\phi}= \frac{\beta}{4} \dot{\phi}^2 + \frac{\phi V(\phi)}{M_{Pl}}$. Taking the differentiation w.r.t. time we have
\begin{equation}
\frac{{\rm d}\rho_{\phi}}{{\rm d}t}= \left[ \frac{\beta}{2}\ddot{\phi} + \frac{1}{M_{Pl}}\left( V(\phi)+ \phi V_{\phi}\right)\right]\dot{\phi}.
\end{equation}
As obvious, the scalar field must satisfy the Klein-Gordon equation (\ref{KG}). Thus using Eq.(\ref{KG}) we have
\begin{equation}
\frac{\beta}{2}\ddot{\phi}= -3H\frac{\beta}{2}\dot{\phi}-\frac{\phi V_{\phi}}{M_{Pl}} -\frac{V(\phi)}{M_{Pl}} + \frac{3V(\phi)}{M_{Pl}}\sqrt{1-\frac{\rho_{\phi}}{\rho_c}}.
\end{equation}
Substituting $\frac{\beta}{2}\ddot{\phi}$ in the expression for $\frac{{\rm d} \rho_{\phi}}{{\rm d}t}$ given above we get,
\begin{equation}
\frac{{\rm d} \rho_{\phi}}{{\rm d}t}= \left[ -3H\frac{\beta}{2}\dot{\phi} + \frac{3V(\phi)}{M_{Pl}} \sqrt{1-\frac{\rho_{\phi}}{\rho_c}}\right] \dot{\phi}.
\end{equation}
As at the present we are looking for deriving the expression for the Raychaudhuri equation of the dynamics dictated by Eq.(\ref{H1}), the Hubble parameter in the above expression must be replaced from Eq.(\ref{H1}) to get 
\begin{eqnarray}
\frac{{\rm d}\rho_{\phi}}{{\rm d}t}&=& \dot{\phi}\left[-3\frac{\beta}{2}\frac{\dot{\phi}}{\phi} \left(\sqrt{\frac{\rho_{\phi}}{3}}-\frac{\dot{\phi}}{2}\right) + 3 \frac{V(\phi)}{M_{Pl}} \right]\sqrt{1-\frac{\rho_{\phi}}{\rho_c}},\\
&=& -3\frac{\dot{\phi}}{\phi} \left[ \frac{\sqrt{3}}{2}\dot{\phi}\sqrt{\rho_{\phi}} -\rho_{\phi} \right]\sqrt{1-\frac{\rho_{\phi}}{\rho_c}},
\end{eqnarray}
where we have used the definition of $\rho_{\phi}$ and the fact that $\beta=3$ to arrive at the above expression. Now substituting $\frac{{\rm d}\rho_{\phi}}{{\rm d}t}$ back into $\frac{{\rm d}H}{{\rm d} t}$, we have
\begin{equation}
\left(\frac{{\rm d}H}{{\rm d} t}\right)_{\rho_{\phi}=\rho_{c}}= \frac{3}{2} \left[ \left( \frac{\dot{\phi}}{\phi^2} \right)\left( \frac{1}{\rho_c}\right)\left(\sqrt{\frac{\rho_{\phi}}{3}} - \frac{\dot{\phi}}{2} \right) \left( \frac{\sqrt{3}}{2} \dot{\phi}\sqrt{\rho_{\phi}}- \rho_{\phi}\right) \right]_{\rho_{\phi}=\rho_{c}}.\label{using}
\end{equation}
Now as $\rho_{\phi}=\frac{\beta}{4}\dot{\phi}^2+\frac{\phi V(\phi)}{M_{Pl}}$, therefore, $\sqrt{\frac{\rho_{\phi}}{3}} = \sqrt{ \frac{\dot{\phi}^2}{4} + \frac{\phi  V(\phi)}{3 M_{Pl}}}$ will always be greater than $\frac{\dot{\phi}}{2}$ for positive valued potential with  $\beta=3$ as in our case. It can be easily seen that $\frac{\sqrt{3}}{2}\dot{\phi}\sqrt{\rho_{\phi}}-\rho_{\phi}$ is always less than zero. Thus we arrive at the conclusion that 
\begin{equation}
\left(\frac{{\rm d}^2 a}{{\rm d}t^2}\right)_{\rho_{\phi}=\rho_c} = \left( \frac{{\rm d}H}{{\rm d}t}\right)_{\rho_{\phi}=\rho_{c}} > 0, ~~{\mbox{only~if}} ~(\dot{\phi})_{\rho_{\phi}=\rho_c}<0.
\end{equation}
Therefore, to have $\rho_{\phi}=\rho_{c}$ as the bouncing point, only the locus of negative initial velocity of the field are allowed. Thus, in this work we consider only the case with $\dot{\phi}_B<0$ as it is the only allowed initial value of the velocity of the scalar field $\phi$ for bounce to happen at $\rho_{\phi}=\rho_{c}$ for the dynamics given by Eq.(\ref{H1}).

Next, we shall examine if $\rho_{\phi}=\rho_{c}$ is also a bouncing solution for Eq.(\ref{H2}).  We begin as above with the identity $\frac{\ddot{a}}{a}= \frac{{\rm d}H}{{\rm d}t} +H^2$ but now with $H$ given by Eq.(\ref{H2}). From Eq.(\ref{H2}) we have seen that the first condition  Eq.(\ref{Condition1}) is satisfied for $\rho_{\phi}=\rho_{c}$ provided that $(\dot{\phi})_{\rho_{\phi}=\rho_c}=0$.  This is a very special case. Now let us examine if this initial condition $(\dot{\phi})_{\rho_{\phi}=\rho_{c}}=0$ also satisfies the second condition for a bounce to occur. This amounts to check if
\begin{equation}
\left( \frac{\ddot{a}}{a} \right)_{\rho_{\phi}=\rho_{c}} = \left(\frac{{\rm d}H}{{\rm d}t}\right)_{\rho_{\phi}=\rho_{c}},
\end{equation}
is greater than zero for $(\dot{\phi})_{\rho_{\phi}=\rho_{c}}=0$. Now $\frac{{\rm d}H}{{\rm d}t}$ for the dynamics given by Eq.(\ref{H2}) is
\begin{equation}
\frac{{\rm d}H}{{\rm d}t}=-\frac{{\rm d}}{{\rm d}t}\left(\frac{\dot{\phi}}{\phi}\right) - \frac{{\rm d}}{{\rm d}t}\left[\frac{1}{\phi} \left(\sqrt{\frac{\rho_{\phi}}{3}} -\frac{\dot{\phi}}{2} \right)\sqrt{1-\frac{\rho_{\phi}}{\rho_c}} \right].
\end{equation}
The second term on the right hand side of the above equation is already evaluated above while analyzing the fulfillment of second bouncing condition for Eq.(\ref{H1}) (for $``+"$ solution of H) and hence using Eq.(\ref{using}) the above equation becomes
\begin{equation}
\frac{{\rm d}H}{{\rm d}t}= -\frac{\ddot{\phi}}{\phi} + \left(\frac{\dot{\phi}}{\phi}\right)^2 + \frac{3}{2} \left( \frac{\dot{\phi}}{\phi} \right)\left( \frac{1}{\rho_c} \right) \left( \sqrt{\frac{\rho_{\phi}}{3}}- \frac{\dot{\phi}}{2}\right) \left( - \frac{\sqrt{3}}{2}\sqrt{\rho_{\phi}} + \rho_{\phi} \right).
\end{equation}
But at $\rho_{\phi}=\rho_{c}$ the initial value of the velocity of the scalar field must be zero to satisfy the first condition of bounce. Therefore, we have 
\begin{equation}
\left(\frac{{\rm d}H}{{\rm d}t}\right)_{\rho_{\phi}=\rho_{c}}= -\left(\frac{\ddot{\phi}}{\phi}\right)_{\rho_{\phi}=\rho_c}.
\end{equation}
Now to check whether the second condition of a bounce is satisfied or not, all we have to do is to evaluate the quantity $-\left(\frac{\ddot{\phi}}{\phi} \right)_{\rho_{\phi}=\rho_c}$ from the Klein-Gordon equation (\ref{KG}) to have,
\begin{equation}
-\frac{\ddot{\phi}}{\phi}= + \frac{3H\dot{\phi}}{\phi} + \frac{2}{\beta M_{Pl}}\frac{\phi V_{\phi}}{\phi} + \frac{2}{\beta M_{Pl}}\frac{V(\phi)}{\phi} -\frac{2}{\beta M_{Pl}} \frac{V(\phi)}{\phi}3\sqrt{1-\frac{\rho_{\phi}}{\rho_{c}}},
\end{equation}
and evaluating the above at $\rho_{\phi}=\rho_c$ we have
\begin{equation}
-\left( \frac{\ddot{\phi}}{\phi}\right)_{\rho_{\phi}=\rho_c}= \frac{2}{\beta M_{Pl}}\left[ V_{\phi} + \frac{V(\phi)}{\phi} \right].
\end{equation}
It can be easily seen that for $\phi$ greater than $M_{Pl}$, we have $\left(\frac{\ddot{a}}{a}\right)_{\rho_{\phi}=\rho_{c}} > 0$, which  gives a bouncing solution for the dynamics given by Eq.(\ref{H2}). Also, tNote that the quantum bounce occurs only for one value of the initial condition given by  $\dot{\phi}_B=0$.  Although this is possible, it is a very special case in  which the kinetic energy at the bounce to be zero.  Since in this paper we are mainly concerned with  the case in which the kinetic energy dominates at the  bounce, we shall not consider this case further. 

Now let us turn to study the background dynamics, for which   we introduce a few parameters.

(1) The Effective equation of state  (EOS), $w(\phi)$: It is defined as the ratio of pressure to energy density 
\begin{equation}
w(\phi)= \frac{P_{\phi}}{\rho_{\phi}}.
\end{equation} 
As we have an effective definition of energy density which is given by $\rho_{\phi}= \frac{\beta}{4}\dot{\phi}^2+ \frac{\phi V(\phi)}{M_{Pl}}$,  an effective definition of pressure of the scalar field $P_{\phi}$ can always be defined if we demand that  $P_{\phi}$ which satisfy the continuity equation 
\begin{equation}
\frac{{\rm d} \rho_{\phi}}{{\rm d}t} + 3H\left(\rho_{\phi}+ P_{\phi}\right)=0.
\end{equation}
Now as the scalar field must satisfy the Klein-Gordon equation, using Eq.(\ref{KG}) an effective expression of pressure can be obtained to be as follows. From the definition of the effective energy density we have
\begin{equation}
\frac{{\rm d} \rho_{\phi}}{{\rm d}t} = \dot{\phi} \left[ \frac{\beta}{2}\ddot{\phi} + \frac{1}{M_{Pl}} \left( V(\phi)+ \phi V_{\phi}\right) \right],
\end{equation}
and from the continuity equation we have 
\begin{eqnarray}
P_{\phi} &=& -\frac{1}{3H}\frac{{\rm d} \rho_{\phi}}{{\rm d} t} - \rho_{\phi}\\ 
         &=& -\frac{1}{3H} \dot{\phi} \left[ \frac{\beta}{2}\ddot{\phi} + \frac{1}{M_{Pl}} \left( V(\phi)+ \phi V_{\phi}\right) \right]- \rho_{\phi}.         
\end{eqnarray}
As the scalar field must satisfy the Klein-Gordon equation, therefore, using Eq.(\ref{KG}) we can write  $\ddot{\phi}$ as,
\begin{eqnarray}
\ddot{\phi} &=& -3H\dot{\phi} - \frac{2}{\beta M_{Pl}}\phi V_{\phi}- \frac{2}{\beta M_{Pl}}V(\phi) + \frac{6}{\beta M_{Pl}} V(\phi)  \sqrt{1- \frac{\rho_{\phi}}{\rho_{c}}}, \\
 \Rightarrow \frac{\beta}{2} \ddot{\phi}  &=& -3H\left(\frac{\beta}{2} \right)\dot{\phi} -\frac{\phi V_{\phi}}{M_{Pl}} - \frac{V(\phi)}{M_{Pl}} + 3\frac{V(\phi)}{M_{Pl}}\sqrt{1-\frac{\rho_{\phi}}{\rho_{c}}}.
\end{eqnarray}
Substituting it into the above equation for $P_{\phi}$ we get
\begin{equation}
P_{\phi}= \frac{\beta}{4}\dot{\phi}^2 - \frac{\phi V(\phi)}{M_{Pl}} - \frac{V(\phi) \dot{\phi}}{H M_{Pl}}\sqrt{1-\frac{\rho_{\phi}}{\rho_c}}.
\end{equation}
Now since we are considering the dynamics of the background given by Eq.(\ref{H1}), let us substitute the Hubble parameter given by Eq.(\ref{H1}) to get explicitly the effective pressure in terms of $\phi$ and $\dot{\phi}$  
\begin{eqnarray}\nonumber
P_{\phi} &=& \frac{\beta}{4} \dot{\phi}^2 - \frac{\phi V(\phi)}{M_{Pl}} - \frac{V(\phi) \dot{\phi} \phi}{M_{Pl} \left(\sqrt{\frac{\rho_{\phi}}{3}} -\frac{\dot{\phi}}{2} \right)}\nb\\
&=& \frac{\beta}{4}\dot{\phi}^2 - \frac{\phi V(\phi)}{M_{Pl}} - 3\dot{\phi} \left[ \sqrt{\frac{1}{3}\left( \frac{\beta}{4}\dot{\phi}^2 + \frac{\phi V(\phi)}{M_{Pl}}\right)}+\frac{\dot{\phi}}{2} \right].
\end{eqnarray} 
Therefore, the effective equation of state parameter $w(\phi)$ is given by 
\begin{equation}
w(\phi) = \frac{P_{\phi}}{\rho_{\phi}} = \frac{\frac{\beta}{4}\dot{\phi}^2 - \frac{\phi V(\phi)}{M_{Pl}} - 3\dot{\phi} \left[ \sqrt{\frac{1}{3}\left( \frac{\beta}{4}\dot{\phi}^2 + \frac{\phi V(\phi)}{M_{Pl}}\right)}+\frac{\dot{\phi}}{2} \right]}{\frac{\beta}{4}\dot{\phi}^2 + \frac{\phi V(\phi)}{M_{Pl}}}.
\end{equation}

(2) The inflation is characterized by acceleration of the universe $\ddot{a}>0$. However it is important to have a prolonged period of inflation to produce enough number of e-folds to cure the problems of Big Bang cosmology. This is often characterized by slow roll parameters. In BD theory with the Starobinsky potential we have three slow roll parameters
\begin{equation}
\epsilon_1 = -\frac{\dot{H}}{H^2}, ~~\epsilon_2 = \frac{\ddot{\phi}}{H\dot{\phi}}, ~~\epsilon_3=\frac{\dot{\phi}}{2H\phi}.
\end{equation}
The slow roll inflation is achieved when $\epsilon_1,~\epsilon_2, \epsilon_3\ll 1$, and it ceases if any  of them becomes 1  \cite{JCAP}.

(3)  The number of e-folds, $N_{inf}$: The amount of expansion of the universe during inflation is characterized as $N_{inf}=ln(\frac{a_f}{a_i})=\int^{t_f}_{t_i} Hdt\simeq \int^{\phi_i}_{\phi_f} \frac{H}{\dot{\phi}} d\phi$. During the phase of the slow roll inflation, using the slow roll condition, this can be readily seen to be
\begin{equation}
N_{inf}= ln\left(\frac{a_f}{a_i} \right) \simeq \int^{\phi_i}_{\phi_{end}} \frac{V}{V'(\phi)} d\phi.
\end{equation}
In this work, we will use the subscript ``i" and ``f" to denote, respectively, the starting and end times of the slow roll inflation. We use $\ddot{a}(t=t_i)=0$ to get the starting time of inflation, whereas  $\epsilon_1$ to obtain the end time of inflation. Though strictly speaking this choice of starting time of inflation is not in accordance with the definition of the slow roll inflation, nevertheless, it is safe to use them because this is insensitive to observations  \cite{o,UnivFeatA}. As the fulfillment of slow roll condition only demands the slow roll parameters to be very very less than one, therefore, there is not a precise convention for the start time. In Ref.\cite{o, UnivFeatA}, it has been shown numerically that the change in the value of $N_{inf}$ due to different choices of start time is $0.2$ for the Starobinsky potential which is negligible for all practical purposes. To support our argument further we refer the reader to \cite{ToSupport}, wherein authors have shown for the end time of inflation the choice of Hubble slow roll parameter or potential slow roll parameter is irrelevant for all practical purposes.

(4) An useful parameter to quantify the dominance of either kinetic energy or potential energy is via $r_{w}\equiv  \frac{\dot{\phi}^2/2}{V(\phi)}$.  It is obvious that for the case of the potential energy dominant  (PED) case at the bounce $r_{w}<1$, whereas it is $r_{w}>1$ for kinetic energy dominant (KED) case. In particular, we are interested in obtaining the critical value $r_w^c$ that would generate 60 e-folds of expansion of the universe in the slow roll regime.

\subsection{Numerical Analysis}\label{NA}

In this subsection we present our numerical results of the background dynamics for a quantum corrected BD cosmology in the Jordan frame for the Starobinsky potential. To study the evolution of the background universe it is sufficient to consider the quantum corrected Friedmann equation (\ref{FD}) and the Klein-Gordon equation (\ref{KG}) as the Raychaudhuri equation can be obtained from these and hence redundant. To begin our numerical evolution we note that the space of the initial data is four dimensional. This includes the initial value of the scale factor $a$, the Hubble parameter $H$, the scalar field $\phi$ and its velocity $\dot{\phi}$.  However the value of the scale factor at the initial time enjoys a constant rescaling freedom without altering the physics. Using this freedom one can fix the scale factor at bounce so that $a_B=1$ without loss of any generality. The choice of the quantum bounce as the initial condition further gives $H=0$ satisfying the first condition  Eq.(\ref{Condition1}) at $\rho_{\phi}=\rho_{c}$.  This in turn determines the initial value of $\dot{\phi}$ up to a sign once we supply the initial value of the scalar field at the bounce $\phi_{B}$.  Hence, with $a_B$ fixed the value of $\phi_B$ and the sign of $\dot{\phi_B}$ completely determine the space of the initial data. In fact the set of initial conditions is the locus of all points on the phase space of $\phi$ and $\dot{\phi}$ satisfying the condition $\rho_c= \frac{\beta}{4}\dot{\phi}_B^2+ V(\phi_B)$. In this paper we concentrate on an initially kinetic energy dominated universe with negative initial sign of $\dot{\phi}.$

In order to simulate the background dynamics we use Eqs.(\ref{KG}) and (\ref{H1}) with initial conditions as defined above. The fact that we choose Eq.(\ref{H1}) over Eq.(\ref{H2}) is because for dynamics dictated by Eq.(\ref{H2}), this is a very special universe which allows only initial conditions satisfying  $\dot{\phi_B}=0$ in order to give a bouncing solution, and this corresponds to an extremely potential dominated universe and therefore is not considered in this work.

The first step to begin the numerical analysis is to calculate the mass of the Starobinsky potential in the Jordan frame, which will be simply adopted from  \cite{JCAP},    $1.4286 \times 10^{-6}~m_{Pl}$. Thus, with the knowledge of the mass we are ready to numerically evolve the system of Eqs.(\ref{KG}) and (\ref{H1}) for a kinetic dominated universe initially. Let us, thus, consider three initial values of $\phi_B/m_{Pl} = ~ (295, 395, 495)$ satisfying the initial bouncing conditions. That is, let us displace the scalar field $\phi$ far away from the minimum of the potential and set the initial velocity of the field to be negative. Note that this negative velocity is required to fulfill the second condition of the bounce as far as dynamics given by Eq.(\ref{H1}) is concerned. Thus, we expect that, if the field is displaced far away from the minimum and set to have a negative velocity initially, it would roll down the potential well, and it would be interesting to see when it will give rise to a slow roll inflation. 

\begin{figure}
$
 \begin{array}{c c}
   \includegraphics[width=0.44\textwidth]{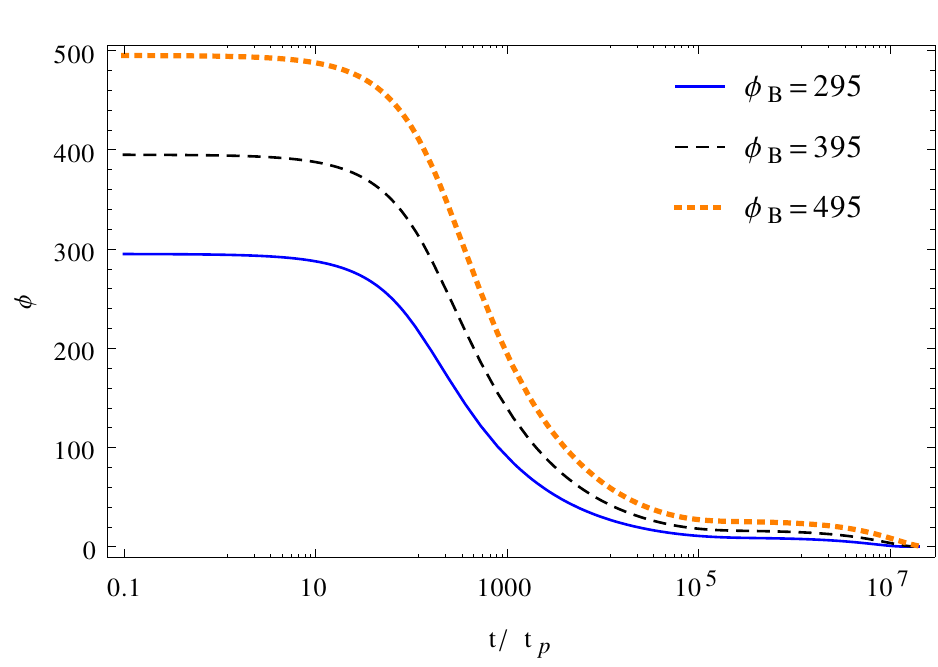} &
   \includegraphics[width=0.45\textwidth]{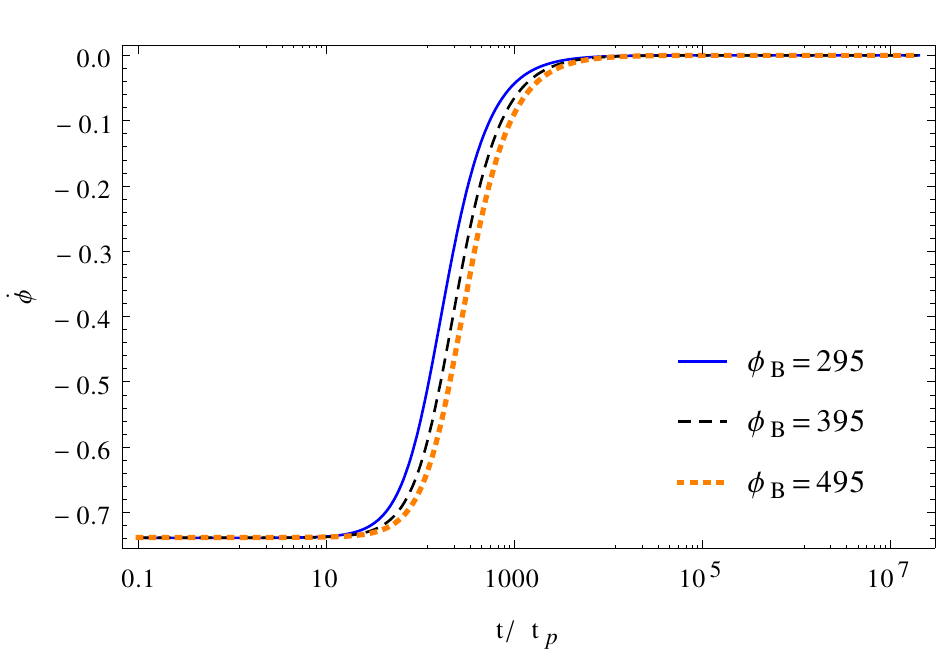} \\
   
 \end{array}
 $
  
  \caption{The \emph{left plot} is for $\phi$ and the \emph{right plot} for $\dot{\phi}$ with  the Starobinsky potential of the form $V(\phi)= \frac{3}{4}M^2 \left(\phi -M_{Pl} \right)^2$ with $M =  1.4286 \times 10^{-6}m_{Pl}$ and  $m_{Pl}=1$ in the Jordan frame for KED case with $\dot{\phi_B}<0$.}
 \label{Fig1} 
\end{figure}

\begin{figure}
$
 \begin{array}{c c}
   \includegraphics[width=0.44\textwidth]{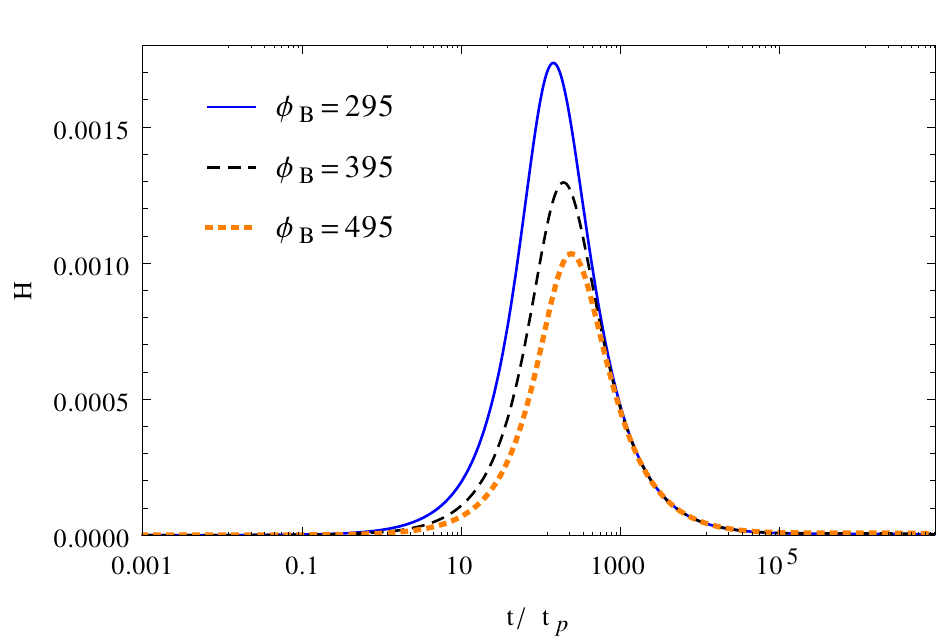} &
   \includegraphics[width=0.454\textwidth]{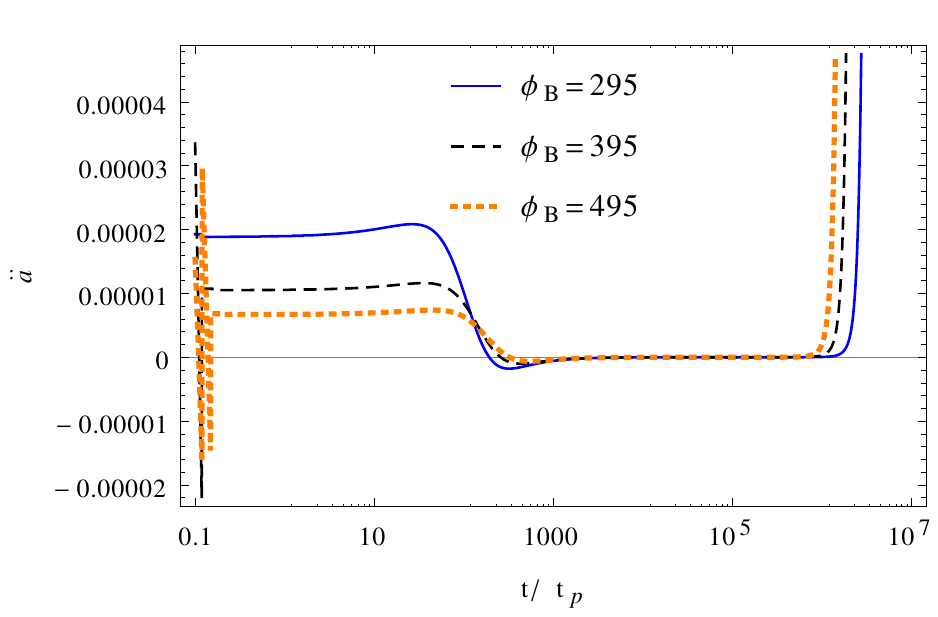} \\
   
 \end{array}
 $
  
  \caption{The \emph{left plot} is for $H$ and the \emph{right plot} for $\ddot{a}$ with  the  Starobinsky potential of the form $V(\phi)= \frac{3}{4}M^2 \left(\phi -M_{Pl} \right)^2$ with $M =  1.4286 \times 10^{-6}m_{Pl}$ and  $m_{Pl}=1$ in the Jordan frame  for KED case with $\dot{\phi_B}<0$.}
 \label{Fig3} 
\end{figure}

The first hint of it can be seen in the plot of Fig.~\ref{Fig1} where we show the variation of the scalar field $\phi$ and its velocity $\dot{\phi}$ for the three initial conditions mentioned above. Note that starting with these initial conditions and with sign of the velocity being negative, eventually, the field $\phi$ will approach the minimum of the potential and correspondingly its velocity becomes zero. Not only that the scalar field approaches the minimum of the potential for all the chosen initial values, but also the variation of the field w.r.t the cosmic time is  negligible after a certain time, say $10^{4}~t/t_{Pl}$,  which is a first clue that the universe might have entered a slowly roll phase of inflation. Before going to analyze whether a slow roll inflation is achieved or not, let us first look at the plot of the Hubble parameter in the lefthand side of Fig.~\ref{Fig3}. It shows that starting from the bouncing point the Hubble parameter $H$ initially increases and then decreases and finally  becomes almost constant. The initial rise of the Hubble parameter indicates a period of super inflation untill it reaches its maximum. Thus, similar to the models of LQC, in BD-LQC it also shows the fact that the quantum geometric effects does not merely solve the initial singularity but its novel repulsive force gives rise to a short period of super inflation  $\frac{{\rm d} H}{{\rm d} t}>0$. 

\begin{figure}
$
 \begin{array}{c c}
   \includegraphics[width=0.44\textwidth]{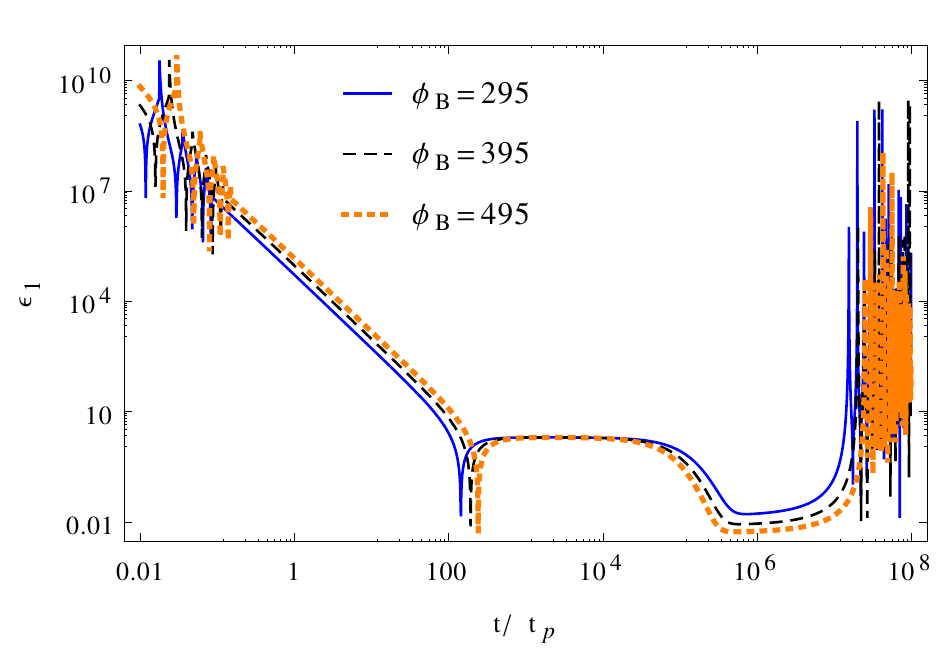} &
   \includegraphics[width=0.429\textwidth]{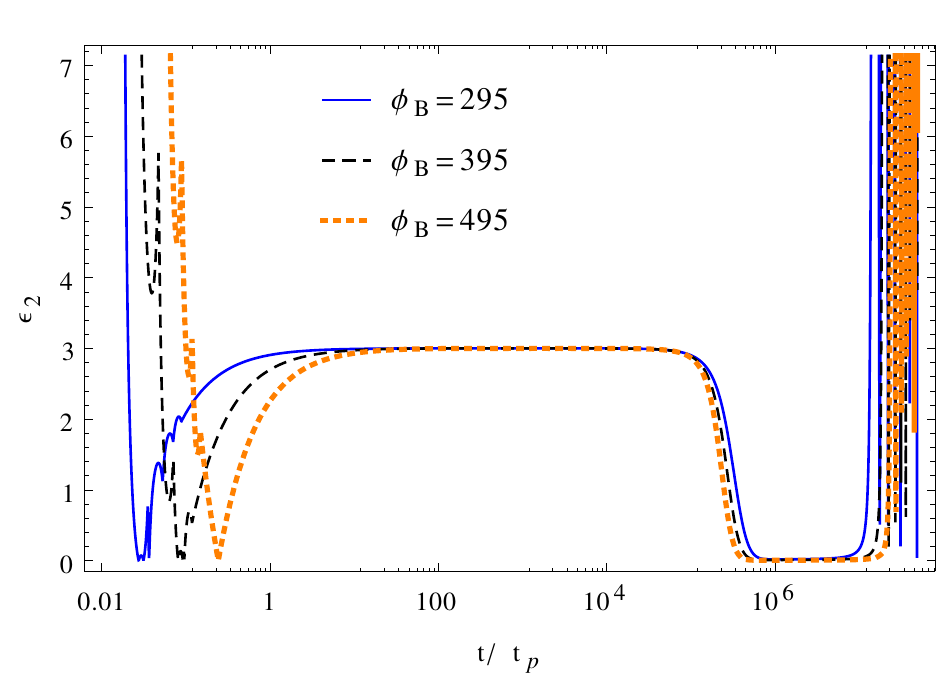} \\
  
 \end{array}
 $
  
  \caption{The \emph{left plot} is for $\epsilon_1$ and the \emph{right plot} for $\epsilon_2$ with  the Starobinsky potential of the form $V(\phi)= \frac{3}{4}M^2 \left(\phi -M_{Pl} \right)^2$ with $M =  1.4286 \times 10^{-6}m_{Pl}$ and $m_{Pl}=1$ in the Jordan frame  for KED case with $\dot{\phi_B}<0$.}
 \label{Fig4} 
\end{figure}

\begin{figure}
$
 \begin{array}{c c}
   \includegraphics[width=0.44\textwidth]{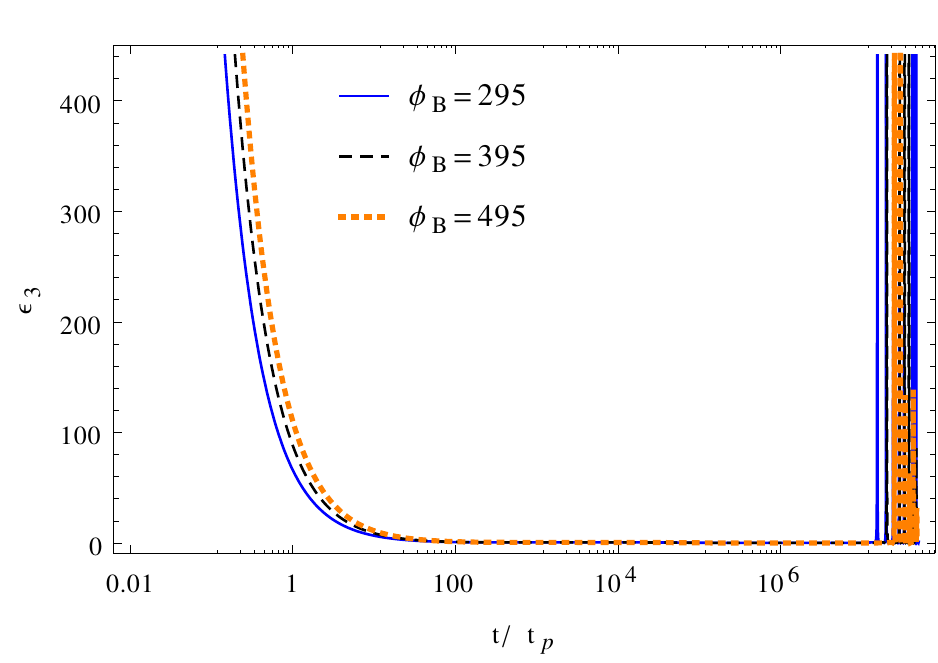} &
   \includegraphics[width=0.463\textwidth]{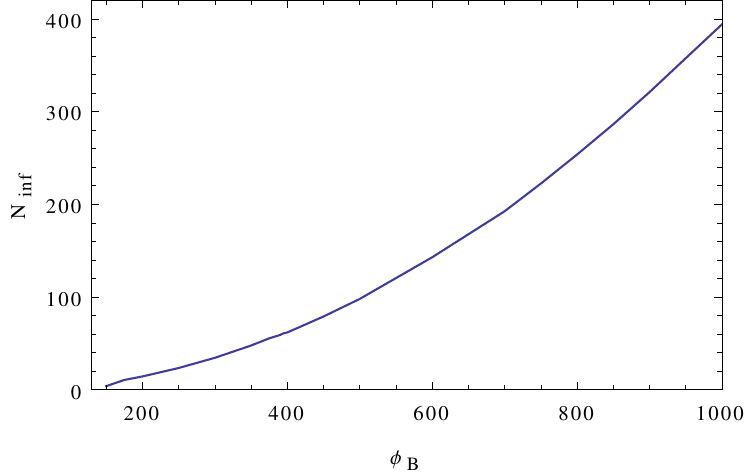} \\
   
 \end{array}
 $
  
  \caption{The \emph{left plot} is for $\epsilon_3$ and the \emph{right plot} for $N_{inf}$ with  the Starobinsky potential of the form $V(\phi)= \frac{3}{4}M^2 \left(\phi -M_{Pl} \right)^2$ with $M =  1.4286 \times 10^{-6}m_{Pl}$ and  $m_{Pl}=1$ in the Jordan frame  for KED case with $\dot{\phi_B}<0$.}
 \label{Fig5} 
\end{figure}

\begin{figure}
$
 \begin{array}{c c}
   \includegraphics[width=0.44\textwidth]{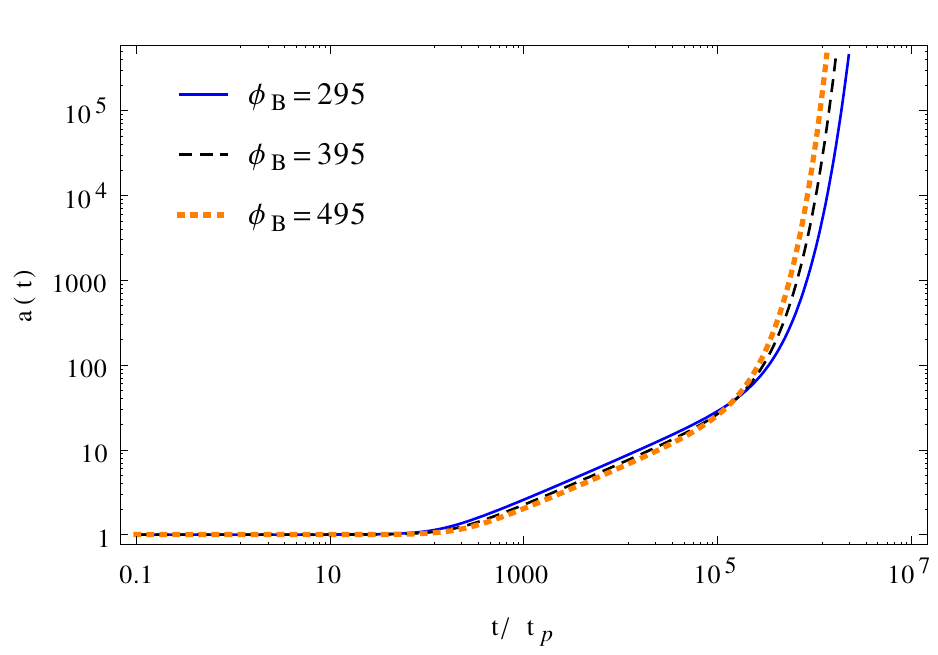} &
   \includegraphics[width=0.43\textwidth]{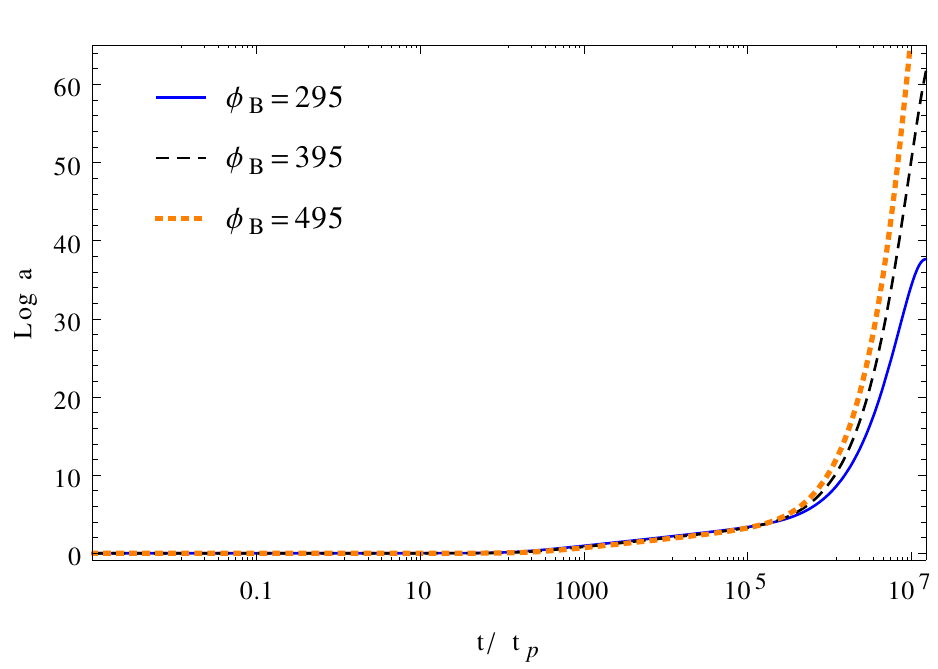} \\
   
 \end{array}
 $
  
  \caption{The \emph{left plot} is for $a$ and the \emph{right plot} for $log~a$ with  the  Starobinsky potential of the form $V(\phi)= \frac{3}{4}M^2 \left(\phi -M_{Pl} \right)^2$ with $M =  1.4286 \times 10^{-6}m_{Pl}$ and  $m_{Pl}=1$ in the Jordan frame  for KED case with $\dot{\phi_B}<0$.}
 \label{Fig2} 
\end{figure}

\begin{figure}
$
 \begin{array}{c c}
   \includegraphics[width=0.44\textwidth]{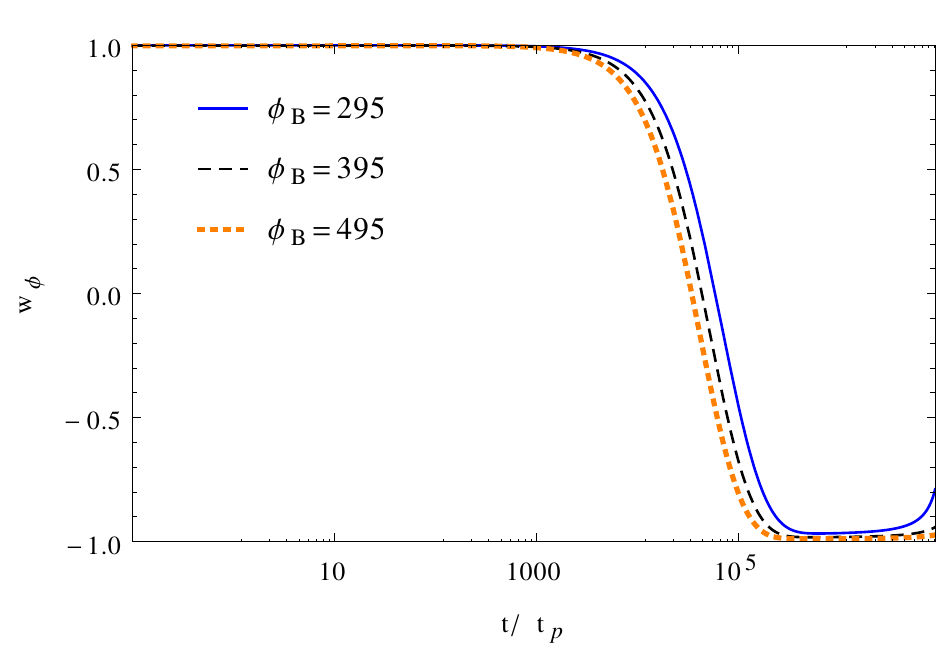} 
 \end{array}
 $
  
  \caption{This plot is for the effective $w_\phi$ of the Starobinsky potential of the form $V(\phi)= \frac{3}{4}M^2 \left(\phi -M_{Pl} \right)^2$ with $M =  1.4286 \times 10^{-6}m_{Pl}$ and $m_{Pl}=1$ in the Jordan frame for KED case with $\dot{\phi_B}<0$.}
 \label{wphi} 
\end{figure}

\begin{figure}
$
 \begin{array}{c c}
   \includegraphics[width=0.44\textwidth]{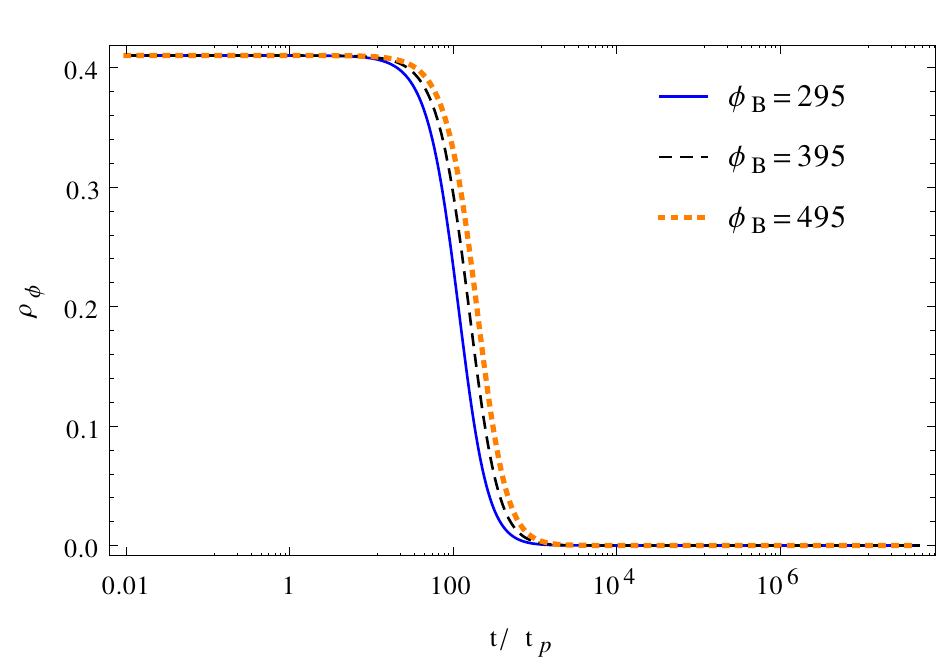} &
   \includegraphics[width=0.463\textwidth]{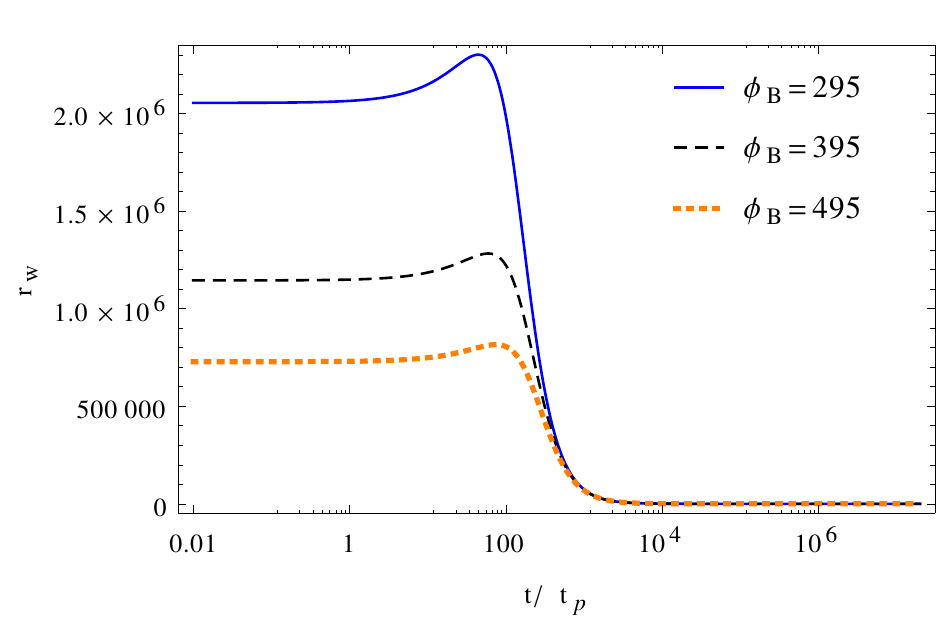} \\
   
 \end{array}
 $
  
  \caption{The \emph{left plot} is for $\rho$ and the \emph{right plot} for $r_w$ with  the Starobinsky potential of the form $V(\phi)= \frac{3}{4}M^2 \left(\phi -M_{Pl} \right)^2$ with $m=  1.4286 \times 10^{-6}m_{Pl}$ and  $m_{Pl}=1$ in the Jordan frame  for KED case with $\dot{\phi_B}<0$.}
 \label{Fig6} 
 \end{figure}

Coming to the plots of the slow roll parameters $\epsilon_1, ~ \epsilon_2, ~ \epsilon_3$, respectively, in Fig.~\ref{Fig4} and in the left-hand side of Fig.~\ref{Fig5}, we can see the fulfillment of    the  slow roll conditions, $\epsilon_1, ~ \epsilon_2, ~ \epsilon_3 \ll 1$ in the period,  $10^4 \sim 10^7$ $t/t_{Pl}$ (for exact values of the beginning and ending  of the inflation see Table \ref{Table1}). Thus a slow roll phase of inflation is achieved. Next, our task is to calculate the number of e-folds of expansion generated during this slow roll phase of inflation  for various values of the initial values of the scalar field $\phi$ and find out what value of $\phi_B$ gives rise to 60 e-folds of inflation. To this purpose we use $\ddot{a}=0$ to obtain the initial time and $\epsilon_1 = 1$ to obtain the end time of inflation. Since by definition $N_{inf}= ln(\frac{a_f}{a_i})$,  therefore, once we have the knowledge of starting and ending times of inflation, it is straintforward to carry out the calculation of $N_{inf}$ from the right plot of Fig.~\ref{Fig2}. Table \ref{Table1} shows $N_{inf}$ as a function of $\phi_B$ for the Starobinsky potential in the Jordan frame for a kinetic energy dominated bounce with $\dot{\phi}_B<0$. Here, we begin with an initial value of $\phi_B=125~m_{Pl}$ to achieve a small number of e-folds of expansion of the order of $\simeq 4$. Since this is not adequate to solve the puzzles of the Standard Model of cosmology, which requires $\sim 60$ e-folds of expansion, we increase in big steps the values of $\phi_B$ and see when $60$ is attained. It is found that $60.83$ number of e-folds are generated for initial condition $\phi_B=395m_{Pl}$ and hence it is considered as the critical value of the initial value we are seeking for and the critical energy density at the bouncing to attain this is $r_w^c = 1.15075 \times 10^6$. Then, we continue this process to investigate the effects of $\phi_{B}$ on $N_{inf}$  up to $\phi_B \simeq 1000 m_{Pl}$ and plot it in the right-hand side of Fig.~\ref{Fig5}. In Fig.~\ref{wphi} we plot the effective equation of state $w_{\phi}$. Starting with a kinetic energy dominated bounce it shows the  three phases of evolution, namely. {\em bouncing phase, transition from quantum bounce to classical universe and the slow roll inflation}. During the bouncing phase as the kinetic energy dominates over potential, the effective equation of state $w_{\phi}\approx1$, while in the slow roll inflation phase $w_{\phi}$ becomes $-1$, which reflects the fact that the  universe is dominated by the potential energy. We also note the universality of the solution in the bouncing phase. Finally in Fig.~\ref{Fig6} we plot the effective energy density due to scalar field $\rho_{\phi}$ on the right-hand side and the ratio   $r_w$ on the left-hand side. The plot of $\rho_{\phi}$ in Fig.~\ref{Fig6} shows the fact that the energy density becomes maximal at the quantum bounce and then dilutes away as the universe expands to large volumes as expected. Whereas  the right-hand side of Fig.~\ref{Fig6} demonstrates the dominance of kinetic over the potential energy throughout the evolution of the universe starting from the quantum bounce. As we initialize the universe with a kinetic energy dominated quantum bounce the ratio $r_w(t_B)\gg1$, and then it transits to $r_w\ll1$ as the universe becomes potential dominated in the slow roll inflation era. 

\begin{figure}
$
 \begin{array}{c c}
   \includegraphics[width=0.438\textwidth]{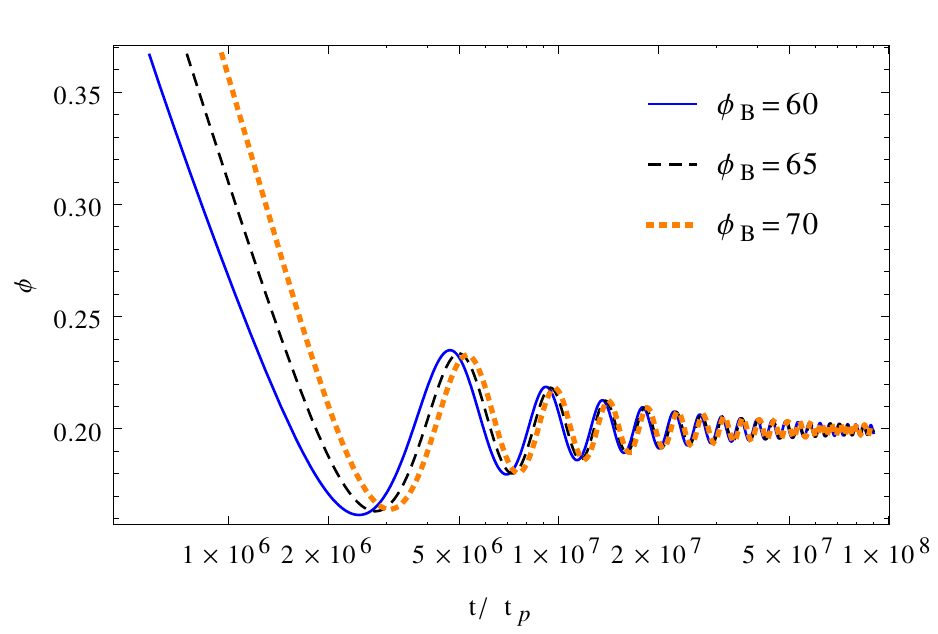} &
   \includegraphics[width=0.463\textwidth]{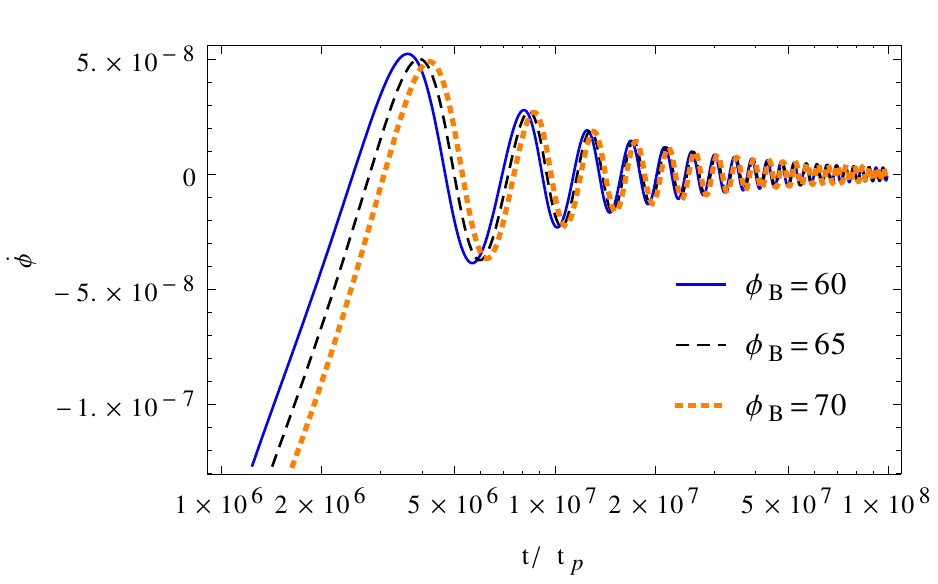} \\
   
   \includegraphics[width=0.463\textwidth]{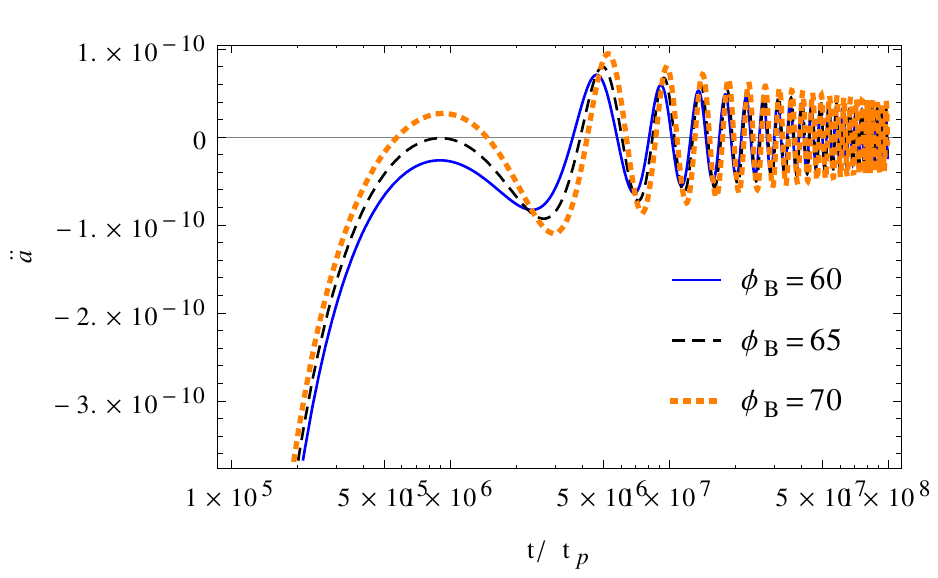} &
   \includegraphics[width=0.41\textwidth]{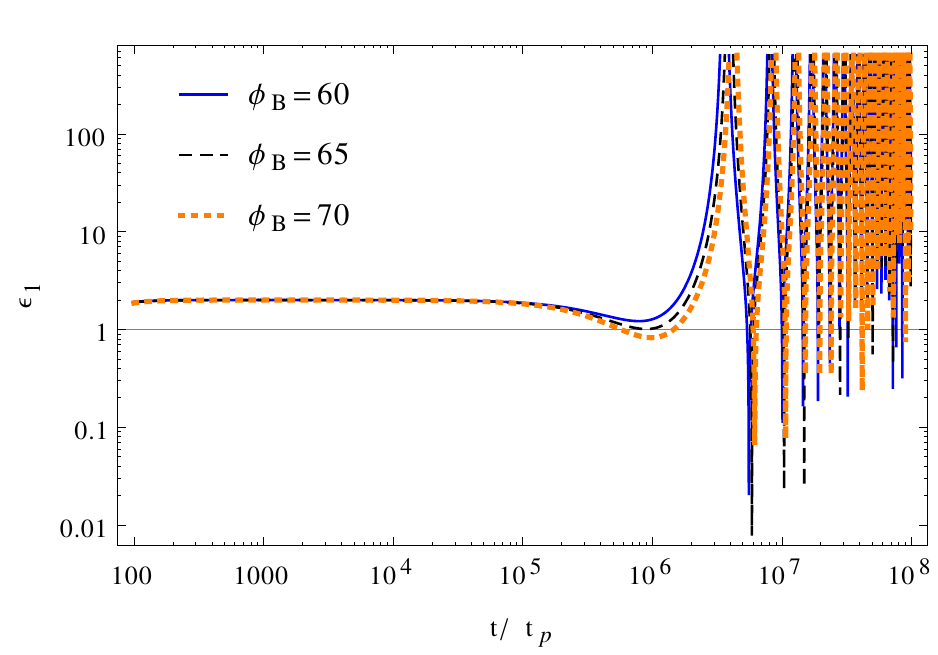} \\
   
 \end{array}
 $
  
  \caption{The \emph{left plot} of top panel  is for $\phi$ and the \emph{right plot} for $\dot{\phi}$ whereas \emph{left plot} of bottom panel is for $\ddot{a}$ and \emph{right} for $\epsilon_1$  with  the Starobinsky potential of the form $V(\phi)= \frac{3}{4}M^2 \left(\phi -M_{Pl} \right)^2$ with $M =  1.4286 \times 10^{-6}m_{Pl}$ and $m_{Pl}=1$ in the Jordan frame  for the KED case with $\dot{\phi_B}<0$ but for those values of $\phi_B$ which do not yield slow roll inflation.}
 \label{Fig7} 
 \end{figure}

We have already observed that a $\phi_B=125 m_{Pl}$ could only produce $4.49$ number of e-folds of expansion of the universe. Thus it would be interesting to see the fate of the universe for lower values of $\phi_B$, say lower than $100 m_{Pl}$.  In Fig.~\ref{Fig7} we analyze this case with $\phi_B=60,~65, ~70~m_{Pl}$ and $\dot{\phi_B}<0$. The evolution of $\phi$ and $\dot{\phi}$ on the top panel of Fig.~\ref{Fig7} suggests that the scalar field and its velocity approach to their respective minimum and start oscillating about it. This is because the inflation quickly rolls down to the minimum of the potential and undergoes damped harmonic oscillations before coming to rest in the asymptotic limit. This damped nature is because of the expansion of the universe as it can be easily seen from the Klein-Gordon equation. Thus for initial values $\phi_B \lesssim 100m_{Pl}$ the universe fails to undergo inflation and hence no slow roll inflation exists. As a proof of it we show the plot of $\ddot{a}$, $\epsilon_1$ on the bottom panel of Fig.\ref{Fig7}. This shows that though the universe shows inflation, i.e. $\ddot{a}>0$, it happens only for a very short period of time as it starts oscillating afterwards. Moreover, a slow roll phase of inflation is never achieved as the Hubble slow roll parameter $\epsilon_1= -\frac{\dot{H}}{H^2}$ is never  less than 1.

\begin{table}
\begin{center}
\caption{This table shows the values of the number of e-folds, $N_{inf}$, for various initial values $\phi_B$  for the potential of the form $V(\phi)= \frac{3}{4}M^2 \left( \phi - M_{Pl}\right)^2$ with  $M=1.4286 \times 10^{-6}m_{Pl}$ in the Jordan frame  for KED case with $\dot{\phi_B}<0$.}
\begin{tabular}{ccccc}
\hline
 $\phi_B$~~~  & Inflation~~~~~~ & $t/t_{pl}$~~~~~~ & $\phi_{*}$~~~~~~ & $N_{inf}$~~~ \\
\hline
 125~~~&starts~~~& $2.19142 \times 10^5$ ~~~& 1.945~~~ & 5.324 \\
 ~~~&ends~~~& $4.41278 \times 10^6$ ~~~& 0.201683~~~ &  \\
 175~~~&starts~~~& $ 1.49064 \times 10^5$ ~~~& 3.91~~~ & 12.1599 \\
 ~~~&ends~~~& $6.92933 \times10^6$ ~~~& 0.257044~~~ &  \\
 200~~~&starts~~~& $1.29018 \times 10^5$ ~~~& 5.179~~~ & 15.0501 \\
 ~~~&ends~~~& $8.15411 \times 10^6$ ~~~& 0.257289~~~ &   \\
 350~~~&starts~~~& $7.1989 \times 10^4$ ~~~& 16.2~~~ & 48.64516 \\
 ~~~&ends~~~& $1.54968 \times 10^7$ ~~~& 0.251808~~~ &  \\
 375~~~&starts~~~& $6.7118 \times 10^4$ ~~~& 18.62~~~ & 53.2791 \\
 ~~~&ends~~~& $1.66544 \times 10^7$ ~~~& 0.2607~~~ &   \\
 390~~~&starts~~~& $6.4485 \times 10^4$ ~~~& 20.08~~~ & 59.984 \\
 ~~~&ends~~~& $1.7398 \times 10^7$ ~~~& 0.256375~~~ & \\
 395~~~&starts~~~& $6.3645 \times 10^4$ ~~~& 20.67~~~ & 61.07 \\
 ~~~&ends~~~& $1.76853 \times 10^7$ ~~~& 0.250834~~~ & \\
 400~~~&starts~~~& $6.2806 \times 10^4$ ~~~& 21.2~~~ & 62.585 \\
 ~~~&ends~~~& $1.7847 \times 10^7$ ~~~& 0.260837~~~ & \\
 500~~~&starts~~~& $5.0076 \times 10^4$ ~~~& 33.19~~~ & 98.475 \\
 ~~~&ends~~~& $2.26954 \times 10^7$ ~~~& 0.2596~~~ & \\
 700~~~&starts~~~& $3.5644 \times 10^4$ ~~~& 65.21~~~ & 193.841 \\
 ~~~&ends~~~& $3.24313\times 10^7$ ~~~& 0.250473~~~ & \\
 800~~~&starts~~~& $3.1144  \times 10^4$ ~~~& 85.2~~~ & 253.527 \\
 ~~~&ends~~~& $3.71644 \times 10^7$ ~~~& 0.2261615~~~ & \\
 850~~~&starts~~~& $2.9280 \times 10^4$ ~~~& 95.33~~~ & 286.21 \\
 ~~~&ends~~~& $3.95935 \times 10^7$ ~~~& 0.2589~~~ &  \\
 900~~~&starts~~~& $2.7618 \times 10^4$ ~~~& 107.6~~~ & 320.829 \\
 ~~~&ends~~~& $4.1991 \times 10^7$ ~~~& 0.259767~~~ & \\
 1000~~~&starts~~~& $2.4848 \times 10^4$ ~~~& 132.7~~~ & 394.839 \\
 ~~~&ends~~~& $4.67784 \times 10^7$ ~~~& 0.26261~~~ & \\
\hline
\end{tabular}
\end{center}
\label{Table1}
\end{table}

\section{Discussion of the result}\label{DR}

This work involves exploring the background dynamics of an effective quantum corrected BD cosmology sourced by a scalar field with the Starobinsky potential in the Jordan frame. Emphasizing on the fact that, though Einstein's frame is equivalent classically to Jordan's frame up to a conformal transformation, it is no longer the case after quantization, we present the classical equations of motion for background dynamics in both  frames. Then, we consider the quantum corrected effective equations of motion in Jordan's frame and analyze in detail whether the bouncing conditions are satisfied when the energy density of the scalar field $\rho_{\phi}$ becomes maximum, that is $\rho_{\phi} = \rho_c$. We find two distinct equations of motion for the Hubble parameter, namely, Eqs.(\ref{H1}) and (\ref{H2}). Though, the Friedmann equation together with the Klein-Gordon equation is sufficient to solve the background dynamics, we  use the Raychaudhuri equation to check whether the second condition of a bounce is satisfied or not. To this effort we derive the corresponding Raycahudhuri equation for each of the Friedmann equations (\ref{H1}) and (\ref{H2}). Thus, analyzing the quantum corrected Raychaudhuri equation we show explicitly that for the universe given by Eq.(\ref{H1}) one can achieve a bounce with  $\dot{\phi}_B<0$. For the dynamics given by Eq.(\ref{H2}) the allowed initial value is too restricted for bounce to occur. In fact, only one initial value is allowed which is the  potential completely-dominated bounce. Since in this paper we are mainly interested in the initially kinetic energy dominant case, this  is not considered.

 After deriving  the Raychaudhuri equation and analyzing the bouncing conditions,  we go on to define an effective equation of state  in the Jordan frame. In order to do so, we first introduce  an effective pressure due to the scalar field. To obtain this, we can always assume that there exists a continuity equation, the same as in LQC of GR. Since we already have the expression of the effective energy density in terms of the scalar field and its velocity $\rho_{\phi}= \frac{\beta}{4}\dot{\phi}^2 + \frac{\phi V(\phi)}{M_{Pl}}$,  and as the scalar field must satisfy the Klein-Gordon equation, therefore, after some simple calculation it is straightforward to obtain an effective pressure.
 

Next we plot the effective equation of state $w_{\phi}$ starting with at the bouncing point. From Fig.\ref{wphi} we notice three phases of the evolution namely. bouncing phase where quantum geometric effects dominate, transition from bouncing to classical phase,  and the slow roll inflation phase. As in the bouncing phase, the kinetic energy dominates  over the potential energy, so we have $w_{\phi} \simeq 1$. Then,  it transits from $+1$ to  $-1$ in the transition phase, and afterward, the universe enters the slow roll phase of inflation when the potential energy dominates. We also note the universality of the solution in the bouncing phase from Fig.\ref{wphi}.

Then, we analyze the slow roll inflation. As one of the main object of this article is to see if starting with an initially bouncing phase one can land up with a slow roll phase of inflation, and if so, for what values of initial conditions we obtain 60 number of e-folds. In order to study the slow roll inflation we choose the dynamics given by Eq.(\ref{H1}) for the Hubble parameter together with the Klein-Gordon equation. The dynamics dictated by Eq.(\ref{H2}) is not considered as it allows only one initial condition for bounce to occur which is the extreme case of potential energy completely-dominated bounce. Also, we note that for dynamics generated by Eq.(\ref{H1}) for any $\phi_B$,  the only allowed sign of $\dot{\phi}_B$ is negative as $\dot{\phi_B}>0$ does not satisfy the second condition of a bounce. And it is obvious that the value of $\dot{\phi}_B$ is fixed by $(\rho_{\phi})_B=\rho_c$  for any given $\phi_B$. This allows us to investigate this interesting situation in which one is free to displace, initially, the scalar field away from the minimum of the potential and setting its initial velocity negative and see if the scalar field rolls down the potential slowly enough to give rise to a slow roll inflation. In BD theory, there are three nontrivial parameters to determine if or not a slow roll phase of inflation is attained. In this work, we obtain the slow roll phase of inflation in quantum corrected BD theory in the Jordan frame for initial values $\phi_{B}\gtrapprox 100 m_{Pl}$. We plot all the relevant cosmological parameters, and show that $\epsilon_1,~\epsilon_2, \epsilon_3 \ll 1$ in the period  $(10^{4} \sim 10^7)t/t_{Pl}$. We also calculate the number of e-folds, $N_{inf}$, generated by various initial conditions at the bounce and dispaly all our results in Table \ref{Table1}. We observe that as we increase $\phi_B$ the number $N_{inf}$  increases and reaches its value $60$ for $\phi_B=395~m_{Pl}$. Finally we plot $N_{inf}$ w.r.t $\phi_B$.

\textbf{Acknowledgement}
A.W. is supported in part by the National Natural Science Foundation of China (NNSCF) with the Grants Nos. 11375153 and 11675145. T.Z. is supported in part by NFSC with Grant No. 11675143 and the Fundamental Research Funds for the Provincial Universities of Zhejiang in China with Grant No. RF-A2019015. M.S. expresses sincere gratitude towards Prof. Yongge Ma for his valuable suggestions and the fruitful discussion.

\end{document}